\documentclass{ar-1col-S2O}

\usepackage[numbers]{natbib}
\usepackage[utf8]{inputenc}
\usepackage{amsmath}
\usepackage{amssymb}
\usepackage{siunitx}
\usepackage{bm}
\usepackage[hidelinks]{hyperref}
\usepackage{mathtools}

\DeclarePairedDelimiterX{\infdivx}[2]{(}{)}{%
  #1\;\delimsize\|\;#2%
}
\newcommand{\infdiv}{\infdivx}

\newcommand{\rmd}{{\rm d}}
\newcommand{\rme}{{\rm e}}
\newcommand{\rmi}{{\rm i}}
\newcommand{\posterior}{p(\bm{\theta} \mid d)}
\newcommand{\prior}{\pi(\bm{\theta})}
\newcommand{\likelihood}{\mathcal{L}(d \mid \bm{\theta})}
\newcommand{\evidence}{\mathcal{Z}(d)}
\newcommand{\inner}[2]{\langle #1 \mid #2 \rangle}
\newcommand{\ncoarse}{{N_{\rm coarse}}}
\newcommand{\los}{\bm{\hat n}}
\newcommand{\tgeo}{t_\oplus}

\newcommand{\chieff}{\chi_{\rm eff}}
\newcommand{\appropto}{\mathrel{\vcenter{
  \offinterlineskip\halign{\hfil$##$\cr
    \propto\cr\noalign{\kern2pt}\sim\cr\noalign{\kern-2pt}}}}}

\newcommand{\removable}[1]{{#1}}

\newcommand{\sk}[1]{}

\jname{Xxxx. Xxx. Xxx. Xxx.}
\jvol{AA}
\jyear{0000}
\doi{10.1146/annurev-nucl-121423-100725}

\begin{document}

% Page header
\markboth{Roulet and Venumadhav}{Inferring Binary Properties from GW Signals}

% Title
\title{Inferring Binary Properties from Gravitational Wave Signals}

% Authors, affiliations address.
\author{Javier Roulet$^1$ and Tejaswi Venumadhav$^{2,3}$
    \affil{$^1$TAPIR, Walter Burke Institute for Theoretical Physics, California Institute of Technology, Pasadena, CA 91125, USA; email: jroulet@caltech.edu}
    \affil{$^2$Department of Physics, University of California at Santa Barbara, Santa Barbara, CA 93106, USA}
    \affil{$^3$International Centre for Theoretical Sciences, Tata Institute of Fundamental Research, Bangalore 560089, India}
}

\begin{abstract}
This review provides a conceptual and technical survey of methods for parameter estimation of gravitational wave signals in ground-based interferometers such as LIGO and Virgo.
We introduce the framework of Bayesian inference and provide an overview of models for the generation and detection of gravitational waves from compact binary mergers, focusing on the essential features that are observable in the signals.
Within the traditional likelihood-based paradigm, we describe various approaches for enhancing the efficiency and robustness of parameter inference.
This includes techniques for accelerating likelihood evaluations, such as heterodyne/relative binning, reduced-order quadrature, multibanding and interpolation.
We also cover methods to simplify the analysis to improve convergence, via reparametrization, importance sampling and marginalization.
We end with a discussion of recent developments in the application of likelihood-free (simulation-based) inference methods to gravitational wave data analysis. 
\end{abstract}

%Keywords, etc.
\maketitle

%Table of Contents
\tableofcontents

\section{INTRODUCTION}

Gravitational waves, predicted by general relativity over 100 years ago, are perturbations in the metric of spacetime sourced by accelerating massive compact objects.
Their existence was first established observationally by indirect means, as responsible for the decay in the orbit of a binary pulsar \cite{Taylor1979}.
In 2015 the Laser Interferometer Gravitational-Wave Observatory (LIGO) made the first direct detection\removable{, sensing the perturbation to the length of the detector due to a passing gravitational wave} \cite{Abbott2016gw150914}.
By now, the LIGO and Virgo interferometers have observed signals from over a hundred mergers of compact binaries hosting black holes or neutron stars, establishing gravitational waves as a major field of research in observational astrophysics \cite{Abbott2023gwtc3, Nitz2023, Olsen2022}.
These detections allow to probe mechanisms of formation and evolution of massive compact binaries over cosmic history, and furthermore enable empirical tests of fundamental questions in physics, such as the equation of state of matter at supranuclear densities and the validity of general relativity in the dynamic, strong-field regime \cite{Christensen2022}.
Extracting these insights requires constructing physical models of the signals and contrasting them with the data to constrain their associated parameters, the subject of this review.

Compared to other astrophysical systems, modeling gravitational waves from binary mergers is, in principle, remarkably clean.
Black holes being the simplest objects in nature, a black-hole binary is completely described with only 17 parameters.
With general relativity we have a complete theory of its dynamics, and the gravitational radiation emitted propagates unobstructed across the universe.
This simplicity makes the problem of parameter estimation well suited for Bayesian inference.
We review all this in Section~\ref{sec:formalism}.
In practice, considerable complications arise.
Modeling waveforms with sufficient accuracy is challenging and requires insights from  perturbation theory in various regimes as well as from expensive numerical simulations.
The signals are weak and their parameters exhibit significant degeneracy, limiting the validity of analytic approximations and calling for sophisticated inference algorithms \cite{Vallisneri2008}.
These can be classified into two broad classes: likelihood-based and likelihood-free, covered in Sections~\ref{sec:likelihood_based} and \ref{sec:likelihood_free}, respectively.

In this review, we focus on methods for parameter estimation in second-generation detectors and the intuition behind them.
We touch briefly on some results only insofar as they help illustrate concepts.
While many of these ideas apply more generally to next-generation ground-based or space-based detectors, we do not discuss those.

\section{FORMALISM}
\label{sec:formalism}

In this section we formulate the problem of Bayesian parameter estimation in mathematical terms, and provide models for the measurement process of gravitational waves and their generation by compact binary coalescences.

\subsection{Bayesian inference}
\label{subsec:inference}

The task at hand is to measure the unknown parameters $\bm{\theta}$ of the physical system, in our case a binary merger, given the data $d$. In the framework of Bayesian inference, this measurement is fully described by the posterior probability density $\posterior$.
By virtue of Bayes' theorem, the posterior can be expressed as
\begin{equation}
    \posterior = \frac{\prior \likelihood}{\evidence}. \label{eq:posterior}
\end{equation}
Here, $\prior$ is the prior probability density, $\likelihood$ is the likelihood of the data, and $\evidence$ is the evidence. 
The prior is the probability density we assign to the parameter values $\bm \theta$ {\em before} the data is analyzed, for us this corresponds to a model of the astrophysical distribution of gravitational wave sources.
The likelihood constitutes a model of the measurement process, including the generation of the signal at the source, the response of the detector to the signal and the noise in the detector data.
The evidence, or partition function, is the normalization constant that makes the posterior integrate to 1 over $\theta$\removable{:
\begin{equation} \label{eq:evidence}
    \evidence = \int \rmd \bm \theta \, \prior \likelihood.
\end{equation}
It is mainly relevant to model selection, which we review in Section~\ref{sec:model_selection}}.

\subsection{Likelihood}
\label{sec:likelihood}

\begin{figure}
    \centering
    \includegraphics{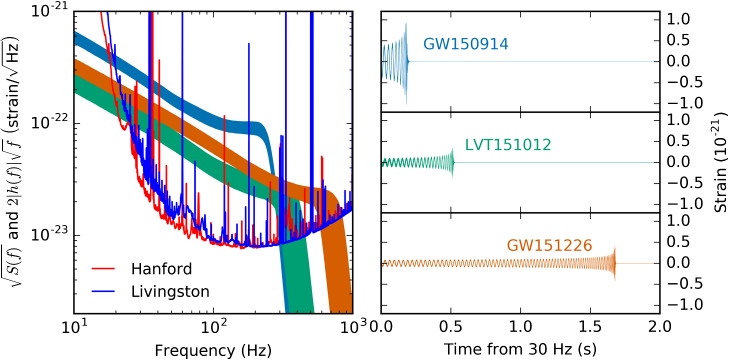}
    \caption{
    \textit{Left panel:} Noise amplitude spectral density $\sqrt{S(f)}$ of the LIGO Hanford and Livingston detectors, and (modified) strain of the first three events detected.
    The detectors are broadband and most sensitive around \SI{100}{\hertz}.
    The signal amplitude scales as $|h| \propto f^{-7/6}$ at low frequencies (inspiral), and terminates at a frequency $\sim M^{-1}$ (merger), see Section~\ref{sec:waveform}.
    \textit{Right panel:} Strains in the time domain. The three systems have decreasing mass from top to bottom.
    Taken from \cite{Abbott2016_o1}.
    % Note: No permission needed since it has CC3.0 license
    % https://journals.aps.org/prx/abstract/10.1103/PhysRevX.6.041015
    }
    \label{fig:o1_events}
\end{figure}

The likelihood is the crucial point for inference via Equation~\ref{eq:posterior}, as it links the data and the parameters. 
We assume that the data is the combination of a signal $h$ and noise $n$, i.e.,
\begin{align}
  d(t) & = h(t) + n(t). \label{eq:additivedata}
\end{align}
We have to solve the inverse problem: start with given data and infer the partition between the signal $h$ and noise $n$. To make progress, we need to know the distribution of the noise $n(t)$. Two common simplifying assumptions are that the noise is stationary and Gaussian in nature, with zero mean. This means that we can completely describe its statistics in terms of its autocorrelation function $C(\tau)$, which is defined by 
\begin{align}
  \langle n(t) \,n(t + \tau) \rangle & = C(\tau).
\end{align}
The $\langle \rangle$ on the left-hand side denotes an expectation value over the distribution that $n$ is drawn from. The right-hand side is a function of only the lag $\tau$ due to the stationary nature of the noise. We define the two-sided power spectral density (PSD) of the noise as the Fourier transform of the autocorrelation:
\begin{equation}
  \label{eq:autocorrtopsd}
  S_2(f) = \int_{-\infty}^\infty \rmd\tau \, C(\tau) \rme^{- \rmi 2\pi f \tau}.
\end{equation}
The autocorrelation function $C(\tau)$ approaches zero in the limit of large lags $\tau$, so the integral in Equation~\ref{eq:autocorrtopsd} converges. The data is real-valued, and hence the autocorrelation $C(\tau)$ and the PSD $S_{2}(f)$ are real-valued and even functions of their arguments. It is convenient to define the one-sided PSD as $S(f) = S_{2} (f) + S_{2} (-f) = 2\, S_{2} (f)$ for frequencies $f > 0$. 
\textbf{Figure~\ref{fig:o1_events}} shows examples of the PSD and strain from three signals detected during the first observing run of Advanced LIGO \cite{Abbott2016_o1}.

We work with finite data segments sampled at an interval $\Delta t$, such that the Nyquist frequency $1/(2 \Delta t)$ encompasses all relevant features of the signal of interest. Given a data segment of duration $T = N \Delta t$, its discrete Fourier transform (DFT) is defined as 
\begin{align}
  \tilde{d}(f_m) & = \Delta t \sum_{n=0}^N\,d(n \Delta t)\,\rme^{-\rmi 2 \pi n \Delta t f_m}, \, {\rm where} \label{eq:dft} \\
  f_m & = \frac{m}{T}, \quad -N/2 + 1 \leq m \leq N/2
\end{align}
If the PSD $S(f)$ is well-behaved and the segment is long, it is approximately true that 
\begin{align}
  \left\langle \tilde{n} \left( f_m \right) \tilde{n}^\ast ( f_{m^\prime} ) \right\rangle \approx \frac{T}{2} \, S\left( f_m \right) \,\delta_{m, m^\prime}, \label{eq:psdlike}
\end{align}
where $\delta_{m, m^\prime}$ is the Kronecker delta function. \sk{In practice, this is never implemented as-is, since the non-periodic nature of the data can lead to artifacts.} Given Equation~\ref{eq:psdlike} for the noise, we can write down the following expression for the likelihood of the data under the hypothesis that the signal has parameters $\bm{\theta}$, called the {\em Whittle likelihood}:
\begin{align}
  \mathcal{L} (d \mid \bm{\theta})
  = \mathcal{L} (n \equiv d - h(\bm{\theta}))
  &\approx \prod_{0 < m < N/2} \frac{2}{\pi T S(f_m)} \exp\left(-\frac{2 \vert \tilde{d} - \tilde{h}(\bm{\theta}) \vert^2}{T S(f_m)}\right). \label{eq:whittle}
\end{align}
We only let the product run over positive frequencies $f_m > 0$, since the negative frequency components are related to them by complex conjugation (the data and noise are real-valued).
To get Equation~\ref{eq:whittle}, we interpreted the real and imaginary parts of the noise Fourier components, $\tilde{n}(f_m)$, as real-valued Gaussian random variables with variance equal to half that of the right-hand side of Equation~\ref{eq:psdlike}.
The zero and Nyquist frequency components are purely real-valued, and including their contribution would change the general form of Equation~\ref{eq:whittle}, so we have omitted them for simplicity. 

The natural logarithm of the likelihood takes the form
\begin{align} 
    \log \mathcal{L}(d \mid \bm{\theta})
    & \approx \sum_{m > 0} \left[ \log{\left( \frac{2}{\pi T S(f_m)} \right)} - \frac{2 \vert \tilde{d} - \tilde{h}(\bm{\theta}) \vert^2}{T S(f_m)} \right] \\
    & \approx
    \big(\text{term independent of }\bm \theta\big)
    - \frac12 \int_0^{f_{\rm max}} \rmd f \, \frac{4 \vert \tilde{d} - \tilde{h}(\bm{\theta}) \vert^2}{S(f)} \\    
    & = \big(\text{term independent of }\bm \theta\big)
    - \frac12 \left \langle d - h(\bm{\theta}) \mid d - h(\bm{\theta}) \right \rangle, \label{eq:loglunsimpl}
\end{align}
where we have defined the inverse-noise-weighted inner product
\begin{equation}
   \left \langle A(t) \mid B(t) \right\rangle
   = 4 \, \Re \int_0^{f_{\rm max}} \rmd f \, \frac{\tilde{A}(f) \tilde B^\ast(f)}{S(f)}. \label{eq:innerproduct}
\end{equation}
We can proceed by expanding the terms in the inner product in Equation~\ref{eq:loglunsimpl}, and collecting those that depend on the signal $h({\bm \theta})$. The other terms constitute the likelihood of the data in the case in which the signal is absent, so we can write
\begin{equation}
   \log \frac{\mathcal{L}(d \mid \bm{\theta})}{\mathcal{L}(d \mid \text{no signal})} = \inner{d}{h(\bm \theta)} - \frac 12 \inner{h(\bm \theta)}{h(\bm \theta)}. \label{eq:gaussian_likelihood}
\end{equation}
Note that the noise within the data is not square integrable, so the integral in the inner product in Equation~\ref{eq:loglunsimpl}, and consequently the log likelihood in Equation~\ref{eq:whittle}, does not approach a finite limit as the duration $T \rightarrow \infty$\removable{\ (this makes sense, as the likelihood is the probability of the data and the data changes as we extend $T$).} The signal $\tilde{h}(f)$, especially when restricted to the sensitive band, is square integrable, so the likelihood ratio of Equation~\ref{eq:gaussian_likelihood} does have a finite limit as $T \rightarrow \infty$.

%Conventionally, we assume that the PSD $S(f)$ is known ahead of time, in which case we omit the terms that are independent of $\bm{\theta}$ and write the log likelihood as
%\begin{align}
%   \log \mathcal{L}(d \mid \bm{\theta}) & = \frac12 \inner{d}{d} +  \inner{d}{h(\theta)} - \frac 12 \inner{h(\theta)}{h(\theta)}, \label{eq:gaussian_likelihood} \\
%   \implies \log \frac{\mathcal{L}(d \mid \bm{\theta})}{\mathcal{L}(d \mid {\rm noise})} & = \inner{d}{h(\theta)} - \frac 12 \inner{h(\theta)}{h(\theta)}
%\end{align}
%where the noise-weighted inner product was defined in Equation~\ref{eq:innerproduct}
%$<\langle d \vert d \rangle$ piece, unknown PSD, determinant term.

% There are several other practical issues we have omitted here, which we will deal with later in Section \ref{sec:realworld}. 

\subsection{Gravitational wave signals}

In this section we give a description of gravitational waves emitted by compact binaries.
We emphasize how various features observable in the signals relate to different physical effects.

\subsubsection{Parameter space}
\label{ssec:parameter_space}
Quasi-circular (i.e., non-eccentric) binary black hole mergers are characterized completely by fifteen parameters: eight ``intrinsic" and seven ``extrinsic". The two component masses and six spin components constitute the intrinsic parameters.
Three extrinsic parameters specify the location of the source relative to Earth (right ascension, declination and distance), three its orientation (inclination, polarization and orbital phase), and one the time of merger.

Eccentric binaries have two additional parameters: the eccentricity and the argument of periapsis.
Gravitational radiation is efficient at dissipating eccentricity \cite{Peters1963}, thus, only binaries that form at close separation and merge promptly can have a measurable eccentricity \cite{Samsing2018}.
In contrast, binaries formed at large orbital separation that later merge due to gravitational wave emission are expected to have quasi-circular orbits.

Finally, if the merger involves neutron stars or other matter objects, additional parameters are required to specify their internal structure.

\subsubsection{Waveform models}
\label{sec:waveform}

Waveform models predict the gravitational wave signal given the parameters of the source.
In principle, this involves solving the dynamics of the compact objects dictated by general relativity (plus matter physics if the objects are not black holes) as they inspiral, merge and ring-down to a single object, and computing the gravitational waves radiated in the process.
This problem has not been solved exactly, but three main approximate methods have been developed over the last decades: numerical relativity, the post-Newtonian expansion, and black hole perturbation theory.
They are complementary to each other since they have different regimes of validity and computational cost.

\textit{Numerical relativity} directly simulates a spacetime and evolves it according to general relativity, it represents the gold standard in terms of accuracy.
On the other hand, it is prohibitively expensive for direct use in parameter estimation, as a single simulation may take months to run.
Numerical relativity is currently applicable for the late inspiral (few tens of orbits), merger and ringdown for moderate mass ratio signals ($q \equiv m_2/m_1 \gtrsim 1/10$).
Multiple groups have supplied catalogs of numerical relativity waveforms \cite{Boyle2019, Healy2020, Ferguson2023,Hamilton2024}.
See \cite{Lehner2014} for a review on numerical relativity.

The \textit{post-Newtonian expansion} is a low-velocity, weak-field approximation in powers of $v/c$.
It is accurate in the early inspiral but breaks down close to merger as the gravitational field and velocity become large.
It features closed-form analytic solutions in both the time and frequency domains.
For the case of aligned-spin systems with quadrupole emission (and using relativist's units $G = c = 1$), the frequency-domain strain signal is given by \cite{Cutler1994, Poisson1995}
{\allowdisplaybreaks
\begin{align}
    h(f) &\appropto Q(\text{angles}) \frac{\mathcal{M}^{5/6}}{D}
        f^{-7/6} \rme^{\rmi \Phi(f)} \label{eq:h} \\
    \Phi(f) &\approx
        %\frac\pi4 +
        2 \phi_0
        - 2 \pi f t_0
        - \frac{3}{128\, \eta v^5} \left[
            1
            + a_2 v^2 + a_3 v^3
            + \mathcal{O}(v^4)
            \right]  \label{eq:pn_phase}\\
    a_2 &= \frac{55}{9}\eta + \frac{3715}{756} \\
    a_3 &= \frac{113}{3} \chieff
        - \frac{38}{3}\eta(\bm\chi_1 + \bm\chi_2)\cdot \bm{\hat{L}}
        - 16 \pi
\end{align}
(we truncated the expansion in $v$, but more terms are known). The parameters above are
\begin{align}
    v &= \left(\pi M f\right)^{1/3} & \text{(orbital velocity)} \label{eq:v}\\
    M &= m_1 + m_2 & \text{(total mass)}\\
    \eta &= \frac{m_1 m_2}{(m_1 + m_2)^2} & \text{(symmetric mass ratio)}\\
    \mathcal M &= \eta^{3/5} M & \text{(chirp mass)}\\
    \chieff &= \frac{m_1 \bm\chi_1 + m_2 \bm\chi_2}{m_1 + m_2} \cdot \bm{\hat{L}} 
        & \text{(effective aligned spin)}\\
    \bm\chi_i &= \frac{\bm S_i}{m_i^2}, \qquad i \in \{1,2\}.
        & \text{(dimensionless spin)}
\end{align}
}
Due to the expansion of the universe, gravitational waves get redshifted as they propagate.
As a result, the measurable quantities are ``redshifted'' \textit{detector-frame} masses $m_{\rm det} = (1 + z) m_{\rm source}$ and luminosity distance $D_L = (1 + z) D_{\rm comoving}$, where $z$ is the redshift \cite{Krolak1987}. 
This is a consequence of the absence of any absolute mass scale in general relativity. 
The effect of cosmological redshift is to rescale the time by $(1+z)$, which is indistinguishable from observing waves from a source with higher mass at a larger distance by the same factor in the absence of cosmological expansion.
The source-frame mass is unobservable, it is inferred indirectly by using a cosmological model to relate the luminosity distance and the redshift.
Thus, the above formulas use detector-frame masses and luminosity distance.
If the source and observer are moving with respect to each other there is an additional Doppler shift and aberration \cite{Bonvin2023}.

\begin{figure}
    \centering
    \includegraphics[width=\linewidth]{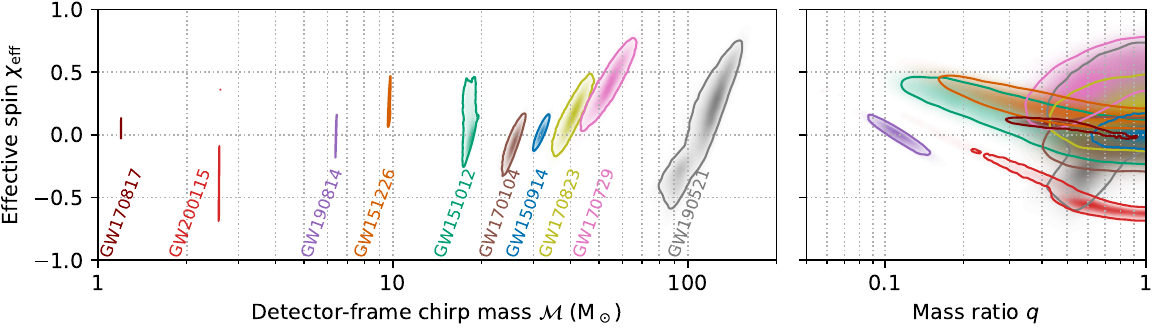}
    \caption{Estimated parameters for an arbitrarily selected subset of signals.
    For low-mass systems, the detector-frame chirp mass is very well measured and the mass ratio and effective spin are degenerate.
    For high-mass events, the mass and effective spin are degenerate.
    (Some posteriors, e.g.\ GW200115, differ from published results \cite{Abbott2023gwtc3}. The reasons are that for this figure we used a broader spin prior, uniform in $\chieff$, and a newer waveform approximant, \texttt{IMRPhenomXODE} \cite{Yu2023}.)}
    \label{fig:events}
\end{figure}

Per Equation~\ref{eq:h}, during the inspiral the amplitude of the waveform scales with $f^{-7/6}$, as can also be seen in \textbf{Figure 1}.
The amplitude profile starts to deviate from this simple power law as the orbital velocity becomes comparable to $c$, at a frequency scale given by the innermost stable circular orbit (ISCO) at which the objects plunge---for a test mass,
$
    f_{\rm ISCO}
    = (6^{3/2} \pi M)^{-1}
    \approx \SI{4.4}{\kilo\hertz}\, (M/\rm M_\odot)^{-1}.
$
After the merger, the emission cuts off rather abruptly as the remnant relaxes to a Kerr black hole.
The ISCO and merger frequencies are affected by the component spins: they increase if spins are aligned with the orbit and decrease if they are anti-aligned, see \cite{Ajith2011} for a fitting function.
Measuring the merger frequency therefore constrains a combination of the total mass and the aligned spins, see the heavier events in the left panel of \textbf{Figure~\ref{fig:events}}.

For the phase, the leading dependence on source parameters is $\propto 1 / (\eta v^5) \propto (\mathcal{M} f)^{-5/3}$, therefore the first quantity we can estimate from the phase evolution is $\mathcal{M}^{-5/3}$.
For low-mass systems, this is large and hence the chirp mass is very well measured, as can also be seen in \textbf{Figure~\ref{fig:events}}.
The next two corrections incrementally involve the mass ratio and aligned spins.
These terms have smaller, and similar, powers of $v$, so they do not evolve so differently over the detector frequency band.
This correlates the measured coefficients, causing a degeneracy between the mass ratio and effective spin which is prominent at low masses \cite{Cutler1994,Poisson1995,Baird2013}.
The degeneracy is exacerbated by the fact that the 1.5 PN coefficient $a_3$ itself combines the mass ratio and spins \cite{Ng2018}.
By Equation~\ref{eq:v}, given a frequency range at which the detector operates, the post-Newtonian expansion is more accurate for lower-mass systems, as they have smaller $v$.

A third approach is black-hole perturbation theory, which treats the metric as a perturbation of a background Schwarzschild or Kerr spacetime \cite{Pound2021}.
For example, the \textit{self-force} approximation is an expansion in the ratio between the curvature scales of the perturbation and background; for binary black holes it becomes an expansion in powers of the mass ratio (which in the limit approaches a test mass in a geodesic orbit).
Other expansions also exist.
While black-hole perturbation theory is much more recent than the other approaches, it is maturing rapidly and starting to inform current models \cite{Islam2022surrogate,Wardell2023,vandeMeent2023}.

Parameter estimation requires a stringent balance of computational efficiency and accuracy for waveform models since state of the art methods require $\sim 10^5$--$10^7$ waveform evaluations for each event. 
For this purpose, different families of \textit{approximants} have been developed, that combine elements of the different approaches mentioned above.
Some approximant families are the \textit{effective one-body} (EOB)---of which two subgroups exist: \texttt{SEOB} \cite{Bohe2017,Cotesta2018,Ossokine2020,RamosBuades2023seobnrv5phm} and \texttt{TEOBResumS} \cite{Nagar2018,Nagar2020,Gamba2022,Nagar2023}---, the \textit{phenomenological} \texttt{IMRPhenom} \cite{Khan2016,Khan2020,GarciaQuiros2020b,Pratten2021,Yu2023,Thompson2024}, and \textit{numerical relativity surrogates} \cite{Blackman2015,Varma2019,Varma2019b}.

\subsubsection{Extrinsic parameters}

The distance $D$, polarization $\psi$, sky location $\los$, geocenter time of arrival $\tgeo$, inclination $\iota$ and orbital phase $\phi$ affect the waveform in an easily computable way.
The gravitational waves radiated by the source in a given direction $(\iota,\phi)$ can be expressed in terms of two polarizations, $h^+(t)$ and $h^\times(t)$.
These, or their Fourier transforms, are the output of a waveform approximant.
The response to this signal measured by a detector $k$ is given by the linear combination
\begin{align} \label{eq:detector_strain}
    h_k(t)
    &= F_k^+(\los, \psi) h^+(t - t_k)
    + F_k^\times(\los, \psi) h^\times(t - t_k), \\
    t_k(\tgeo, \los) &= \tgeo - \los \cdot \bm r_k/c.
\end{align}
The antenna responses $F^+, F^\times$ have known forms in terms of trigonometric functions \cite{Whelan2013}. 
The arrival time at the detector, $t_k$, depends on its location $\bm r_k$ due to the finite speed of gravitational waves $c$. 
As we will explain in more detail in Section \ref{sec:higher_modes} the inclination determines the polarization content of the wave; for aligned-spin, quadrupole waveforms Equation~\ref{eq:detector_strain} reduces to
\begin{equation}
    \tilde h_k(f)
    = \left[\frac{1+\cos^2\iota}{2} F_k^+(\los, \psi)
    - \rmi \cos \iota F_k^\times(\los, \psi)\right]
    \rme^{-\rmi 2 \pi f t_k(\tgeo, \los)} \rme^{\rmi 2 \phi} \frac 1D \tilde h_0(f),
    \label{eq:22aligned}
\end{equation}
where $\tilde h_0$ only depends on intrinsic parameters.
In particular, waves from face-on or -off sources ($\cos\iota = \pm 1$) are circularly polarized and those from edge-on sources ($\cos\iota = 0$), linearly polarized.

\subsubsection{Spin--orbit precession}

If the spins of the compact objects are aligned with the orbital angular momentum, the system has a reflection symmetry about the orbital plane.
The evolution preserves this symmetry and hence the orbital plane stays fixed.
If the spins are misaligned, this symmetry is broken and the spin and orbital angular momenta exhibit precession.
For most configurations, the direction of the total angular momentum $\bm J$ is approximately conserved and $\bm L, \bm S_1, \bm S_2$ precess around it \cite{Apostolatos1994}.
This is because over the course of a precession cycle, the components of the radiated angular momentum that are perpendicular to $\bm J$ approximately average out, thus $\bm J$ decreases in magnitude but has a fixed direction.
In general the two spins precess at different rates, causing an additional nutation of $\bm L$ \cite{Gangardt2021}.
This picture is justified early in the inspiral, where there is a clean separation of timescales: orbits are much faster than precession cycles, which in turn are much faster than the dissipation timescale at which the system loses energy to gravitational waves.
As the binary approaches merger, these timescales become comparable to each other and are of order the mass of the remnant.
The behavior just described can break if at some point the spin approximately cancels the orbital angular momentum, for example, for a binary with unequal masses (thus, small $L$) and large anti-aligned primary spin. Then, $J \ll L$ and a small change to $\bm J$ significantly affects its direction.
In this case the binary tumbles, a process known as \textit{transitional precession} \cite{Apostolatos1994}.
Richer spin resonances can also arise from the interplay between precession and nutation \cite{Gangardt2021}.

Phenomenologically, compared to aligned-spin systems, precession introduces amplitude modulations in the waveform, along with corrections to the phase evolution (both secular and oscillatory).
One way to understand these effects is by introducing the non-inertial co-precessing frame, which rotates in such a way that the orbital angular momentum direction $\bm {\hat L}$ appears fixed \cite{Schmidt2011}.
In this frame, the gravitational wave emission resembles that of an aligned-spin system---with some late-inspiral corrections to the orbital phase due to spin--orbit and spin--spin coupling, which involve the in-plane spins.
The co-precessing and inertial frames are related by a time-dependent rotation at the (slow) precessional timescale.
Per Equation~\ref{eq:22aligned}, a binary emits gravitational waves preferentially perpendicular to the orbital plane.
Depending on the location of the observer relative to the source, the direction of maximum emission may alternatively point towards or away from the observer as the orbital plane wobbles.
When this happens, the observed amplitude of the wave shows modulations on the precession timescale. 

The effects of precession on the waveform can be described equivalently in terms of \emph{precession harmonics} \cite{Fairhurst2020}.
For a given radiation multipole $(\ell, m')$ in the co-precessing frame, the instantaneous rotation connecting to the inertial frame will leave $\ell$ invariant and mix the $m$, manifesting as $2\ell + 1$ harmonics with $-\ell \leq m \leq \ell$.
In particular, the dominant co-precessing mode $(\ell, m') = (2, 2)$ yields five harmonics in the inertial frame.
Each precession harmonic receives a factor $\rme^{\rmi m \alpha(t)}$ from the time-dependent rotation, where $\alpha$ is the precession angle, so their frequencies differ by the precession frequency $\Omega_{\rm p} = \dot \alpha$.
In this picture, the amplitude modulations arise from the beating of the harmonics.
The harmonics form a power series in $\tan(\beta/2)$, where $\beta$ is the angle between $\bm J$ and $\bm L$.
For most configurations $\beta$ is small, and then the leading effect of precession is well approximated by the first two harmonics.

\subsubsection{Higher-order multipoles}
\label{sec:higher_modes}

The gravitational radiation emitted by a localized source can be decomposed in multipoles that govern the angular dependency of the wave in the far-field \cite{Thorne1980}.
Since the metric perturbation is a tensor, the usual (scalar) spherical harmonics are not adequate to describe it.
We say that a function ${}_sf(\theta, \phi)$ on the sphere has spin $s$ if it transforms according to ${}_sf'(\theta, \phi) = {}_sf(\theta, \phi) \rme^{-\rmi s \varphi}$ under a rotation by $\varphi$ about $\los(\theta, \phi)$.
The spin-weighted spherical harmonics ${}_sY_{\ell m}(\theta, \phi)$ are an orthonormal basis for such functions.
We can construct a quantity with spin $-2$ out of the gravitational wave polarizations as $h_+ - \rmi h_\times$, and decompose it using spin-weighted spherical harmonics with  $s=-2$:
\begin{equation} \label{eq:hm_decomposition}
    h_+ - \rmi h_\times
    = \sum_{\ell \geq 2} \sum_{m=-\ell}^\ell
        {}_{-2}Y_{\ell m}(\iota, \phi) h_{\ell m}.
\end{equation}
Expressions for ${}_{-2}Y_{\ell m}$ and $h_{\ell m}$ can be found e.g.\ in \cite{Blanchet2014}; here we will simply point out a few relevant properties and special situations where some modes vanish.
For this discussion we assume non-precessing binaries (or, alternatively, use the co-precessing frame), and we align the $z$-axis with the orbital angular momentum.

Gravitational radiation has $\ell \geq 2$: the $\ell = 0$ (monopole) and $\ell = 1$ (dipole) multipoles are absent due to conservation of energy and momentum, respectively. 
%\teja{There are edge cases where these things can happen in the near field I think... https://arxiv.org/abs/1002.0351, https://academic.oup.com/mnras/article/345/1/L1/984974... in the far field, you can't do this and mathematically, the appropriate ${}_{-2}Y_{\ell m}$ don't exist. Maybe we can emphasize that the expansion in Eq.~\eqref{eq:hm_decomposition} is a far-field thing?}.
Inspiraling binaries emit dominantly in the quadrupole mode $(\ell, |m|) = (2, 2)$, although other multipoles---usually referred to as \textit{higher modes} or \textit{higher harmonics}---are generically present and have in fact been detected in some of the signals \cite{Abbott2020_GW190412,Abbott2020_GW190814}.
Depending on the source parameters and detector frequency response, the next modes in importance are usually $(3, 3)$ and $(4, 4)$.

Just like scalar harmonics, spin-weighted harmonics are of the form
\begin{equation}
    {}_{s}Y_{\ell m}(\iota, \phi)
    = {}_{s}A_{\ell m}(\iota) \rme^{\rmi m \phi},
    \qquad {}_{s}A_{\ell m} \in \mathbb{R}.
\end{equation}
All ${}_{-2}Y_{\ell m}$ with $|m|\neq2$ vanish at the poles; this can be understood since the metric perturbation must transform with $\rme^{\rmi 2 \varphi}$ under a rotation by $\varphi$, which matches $\phi$ or $-\phi$ at the two poles.
Moreover, modes $h_{\ell m}$ with odd $m$ vanish for equal-mass binaries, as in that case an azimuthal rotation by $\pi$ becomes a symmetry and only modes with $\rme^{\rmi m \pi} = 1$ survive.
Hence, the most favorable systems for observing higher modes are inclined (edge-on), unequal-mass binaries.

Another useful property is that, during the inspiral, a mode $(\ell, m)$ undergoes $m$ cycles per orbit: its phase is approximately $\Phi_{\ell m}(t) \approx m \,\Phi_{\rm orb}(t)$ or, in the frequency domain \cite{GarciaQuiros2020b},
\begin{equation} \label{eq:hm_phase}
    \Phi_{\ell m}(f) \approx m \,\Phi_{\rm orb}\left(\frac fm\right).
\end{equation}
Thus, in general higher modes have higher frequencies, which can affect their detectability due to the uneven detector response to different frequencies. For high-mass systems, which merge at a low orbital frequency, the quadrupole may fall largely outside the detector band, while the higher modes do not get as suppressed \cite{Mills2021}.
The simple scaling in Equation~\ref{eq:hm_phase} breaks as the binary merges and relaxes, or \textit{rings down}, to a Kerr hole.
The ringdown can be described in terms of a superposition of \textit{quasinormal modes} with characteristic frequencies, damping times and angular patterns of radiation.
These frequencies are determined by the mass and spin of the remnant, and depend strongly on $\ell$ as well as $m$ \cite{Berti2006}.

\begin{figure}
    \centering
    \includegraphics[width=.5\linewidth]{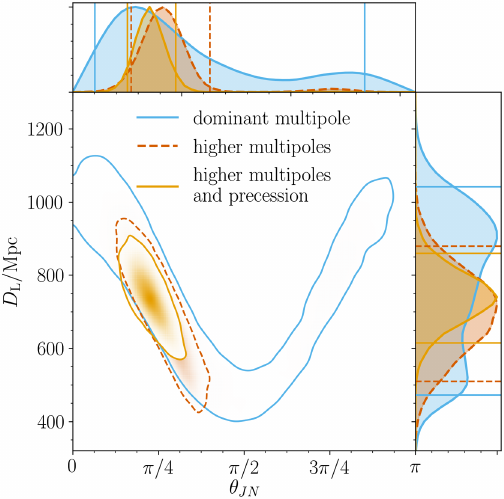}
    \caption{Posterior for the inclination and distance of GW190412, under increasingly refined waveform models. As the effects of higher multipoles and precession are sequentially added, the uncertainty on the inclination, and thereby on the distance, gets reduced.
    Taken from \cite{Abbott2020_GW190412}.
    % No permission needed since it has CC4.0 license 
    % https://journals.aps.org/prd/abstract/10.1103/PhysRevD.102.043015
    }
    \label{fig:GW190412_inclination}
\end{figure}

In the context of parameter estimation, there is good motivation to model higher modes.
By virtue of Equation~\ref{eq:hm_decomposition}, when multiple modes are present the signal depends in a nuanced way on the inclination and phase (via ${}_{-2}Y_{\ell m}$) as well as intrinsic parameters (via $h_{\ell m}$).
Notably, quadrupole-only waveforms exhibit degeneracies between the effective spin and mass ratio, and between the distance and inclination.
The relative amplitude of the higher harmonics depends explicitly on the mass ratio and the inclination, breaking these degeneracies and allowing to measure those parameters separately.
This is illustrated in \textbf{Figure~\ref{fig:GW190412_inclination}} for the event GW190412 \cite{Abbott2020_GW190412}: including higher modes (and precession) in the model helps constrain the inclination and in turn the luminosity distance, which is otherwise degenerate.
Constraining the mass ratio allows to measure the component masses individually beyond the chirp mass, for example, the secondary object in GW190814 has a remarkable mass $m_2 = 2.59^{+0.09}_{-0.08}\,\rm M_\odot$, heavier than any neutron star and lighter than any black hole otherwise measured \cite{Abbott2020_GW190814}.
Improving the distance constraints can potentially aid efforts to search for electromagnetic counterparts by ruling out certain galaxies as possible hosts.
With future detector networks, it may even allow to single out the host galaxy without the need for an electromagnetic counterpart, enabling among other things accurate measurement of the Hubble parameter \cite{Borhanian2020}.

\removable{For precessing systems, we distinguish between co-precessing frame and inertial frame $m$ numbers.
In both cases the waveform is measured in the inertial frame, but the decomposition into modes is different, with different desirable properties:
modes with the same co-precessing frame $m$ have a smooth amplitude and phase obeying Equation~\ref{eq:hm_phase}, while the inertial frame $(\ell, m)$ determine how the waveform transforms under rotations per Equation~\ref{eq:hm_decomposition}.
The two frames are related by a slowly-varying (in the precession timescale) 3D rotation.
The $\ell$ number is invariant under rotations so, in the adiabatic regime, it is the same in both frames.
For non-precessing (aligned-spin) signals, the two frames coincide.
}

\subsubsection{Eccentricity}

As mentioned earlier, eccentricity gets dissipated during the inspiral, moreover, in general relativity it is challenging to even define eccentricity because it is gauge dependent.
As a result there are multiple definitions in use that satisfy the appropriate Keplerian limit \cite{Loutrel2018}.
One practical definition is in terms of the waveform at infinity \cite{Mora2004}, see \cite{Shaikh2023} for an open-source implementation.
Phenomenologically, eccentricity excites emission in all harmonics of the orbital frequency.
Modeling eccentric systems is a topic of active research: recent developments include the incorporation of higher harmonics \cite{Liu2022,RamosBuades2023} and misaligned spins \cite{Liu2023}.

\subsubsection{Tides}
If one or both of the merging compact objects are not black holes (e.g., neutron stars) the orbital dynamics are affected.
Each star is distorted by the tidal field from the companion, extracting energy from the orbit, and additionally inducing a quadrupole moment that adds to that of the binary itself, enhancing gravitational wave emission.
These two effects are similar in importance, and both tend to accelerate the inspiral, especially towards the end as the separation between the objects gets comparable to their size.

To linear order, the induced quadrupole is proportional to the tidal field, $Q_{ij} = -\lambda \varepsilon_{ij}$.
The proportionality constant $\lambda$ is called tidal deformability, and depends on the equation of state and mass of the star.
A dimensionless deformability can be constructed as $\Lambda = \lambda / m^5 \sim (R/m)^5$, where $R$ and $m$ are the star's radius and mass.
The leading-order change to the waveform phase in Equation~\ref{eq:pn_phase} is only at 5th post-Newtonian order: $\Delta\Phi_{\rm tidal}(f) = (117/256\eta)\tilde\Lambda v^5(f)$, where the parameter $\tilde \Lambda$ is a combination of the component masses and deformabilities, which for identical objects reduces to $\Lambda_1 = \Lambda_2 = \tilde \Lambda$ \cite{Flanagan2008, Favata2014}.
Although the phase correction is suppressed by a high power of $v/c$, its prefactor is large for neutron stars because $(R/m)^5 \gg 1$, rendering this effect potentially measurable.
The dimensionless deformability is larger, thus easier to measure, for less massive neutron stars. 
The binary neutron star merger GW170817 was constrained to have $\tilde \Lambda \lesssim 800$ \cite{Abbott2017_GW170817}, which together with the chirp mass determination informs the neutron stars' equation of state and radius \cite{Abbott2018_GW170817_EOS, Raithel2018}.
See \cite{Chatziioannou2020} for a detailed review.

\subsection{Model selection}
\label{sec:model_selection}

Sometimes we may have have multiple competing models, or hypotheses $\mathcal H_1, \mathcal H_2, \ldots$, to interpret the data.
For example, we may consider different distributions of black hole spins (encoded in the prior), or different theories of gravity (encoded in the likelihood).
Bayesian statistics offers a way to assign a ``posterior probability'' to the models themselves in order to quantify how well they explain the observations.
In this context the Bayesian evidence $\mathcal Z(d \mid \mathcal H)$ is nothing but the likelihood for the whole model $\mathcal H$ (implicit in Equation~\ref{eq:evidence} in the form of the prior and likelihood), aggregated over the range of predictions $\mathcal H$ makes by marginalizing over its internal parameters $\bm \theta$.
For any two models, the ratio of these probabilities is the \textit{odds}
\begin{equation}
    \mathcal O^1_2
    \equiv \frac{p(\mathcal H_1 \mid d)}
           {p(\mathcal H_2 \mid d)}
    = \frac{\pi(\mathcal H_1)}
           {\pi(\mathcal H_2)} \cdot
        \frac{\mathcal Z(d \mid \mathcal H_1)}
             {\mathcal Z(d \mid \mathcal H_2)}
    = \frac{\pi(\mathcal H_1)}
           {\pi(\mathcal H_2)} \cdot
        \frac{\int \rmd \bm \theta_1 \, \pi(\bm \theta_1 \mid \mathcal H_1)
              \mathcal L(d \mid \bm \theta_1, \mathcal H_1)}
             {\int \rmd \bm \theta_2 \, \pi(\bm \theta_2 \mid \mathcal H_2)
              \mathcal L(d \mid \bm \theta_2, \mathcal H_2)}.
\end{equation}
If $\mathcal O^1_2 \gg 1$ we say that model $\mathcal H_1$ is favored over $\mathcal H_2$ and vice-versa.
Note that the parameter spaces of the two models need not be the same.
The part of the odds that depends on the data is the ratio of evidences, called Bayes factor between the two models.
As a practical note, model selection is one application where it is important that the normalization is consistent across all the models compared.
Another caveat to keep in mind is that the odds do not provide a goodness-of-fit test: both models might be bad at describing the data even if the odds favor one over the other.
See \cite{Thrane2019} for a review with emphasis on Bayesian inference.
Algorithms for computing the evidence $\mathcal Z$ are comprehensively reviewed in \cite{Christensen2022}.

\subsection{Representation of the posterior}
\label{sec:representation}
As discussed in Section~\ref{ssec:parameter_space}, binary mergers are characterized by $\gtrsim 15$ parameters.
This dimensionality is prohibitively large for the straightforward approach of tiling the parameter space with a dense grid and recording the posterior probability everywhere.

One efficient way of describing the high-dimensional posterior distribution is by reporting samples drawn randomly from it, $\{\bm \theta_i\} \sim \posterior$.
Samples from the distribution are very versatile, because they allow Monte Carlo estimation of integrals of the form
\begin{equation}
        \int \rmd \bm \theta\, p(\bm \theta) f(\bm \theta) \approx \frac 1 N \sum_{\bm \theta_i \sim p}^N f(\bm \theta_i).
    \label{eq:montecarlo_integral}
\end{equation}
for arbitrary functions $f$.
A simple example is a histogram of the samples to estimate a marginal distribution: the counts in each bin can be expressed in the form \ref{eq:montecarlo_integral} by choosing $f(\bm \theta) = 1$ inside the bin and 0 outside.
Other summary statistics like the mean or quantiles of the distribution are expressible in this form as well.
Such integrals also arise in population inference, for evaluating the likelihood of models of the astrophysical distribution.

A second way of representing the posterior is by fitting a parametric functional form.
For gravitational-wave applications the posterior usually has a complicated structure, motivating the use of flexible approaches such as mixture models \cite{Chua2020,Gabbard2021} or normalizing flows \cite{Green2020,Green2021gw150914,Dax2021}, see Section~\ref{sec:likelihood_free}.
Those models also allow to generate samples easily, so the two methods can be combined.

\section{LIKELIHOOD-BASED METHODS}
\label{sec:likelihood_based}

A large class of parameter estimation methods rely on the ability to evaluate the likelihood function given a choice of parameters.
This requires an explicit model for the noise, which nearly always is assumed stationary and Gaussian, leading to the Whittle likelihood \ref{eq:gaussian_likelihood}.

General-purpose algorithms can generate samples from a distribution given the ability to evaluate the probability density at arbitrary points.
These samplers explore the parameter space efficiently by using the result of previous evaluations to guide new proposals.
Two such algorithms frequently used in gravitational wave inference are nested sampling \cite{Skilling2006} and Markov Chain Monte Carlo (MCMC) \cite{Metropolis1953, Hastings1970}, overviewed e.g.\ in \cite{Christensen2022,Veitch2015}.

While sampling algorithms make parameter estimation tractable, in general the process is computationally expensive and not always robust.
In this section we review several methods developed to improve the speed and convergence of sampling algorithms in the specific context of gravitational waves.
In likelihood-based parameter estimation, most of the computational effort is generally spent evaluating the likelihood.
To improve this, some techniques aim at making each individual likelihood evaluation faster, and others at decreasing the number of likelihood evaluations required for convergence.
We review these approaches in Sections \ref{sec:likelihood_acceleration} and \ref{sec:degeneracy}, respectively.
Finally, other methods try to simplify the sampled distribution by  correcting (reweighting) samples taken from an approximating distribution (Section~\ref{sec:importance_sampling}) or marginalizing some parameters (Section~\ref{sec:marginalization}).

Codes for likelihood-based inference of gravitational wave sources include
\texttt{LALInference} \cite{Veitch2015},
\texttt{BayesWave} \cite{Cornish2015},
\texttt{BAYESTAR} \cite{Singer2016},
\texttt{RIFT} \cite{Lange2018},
\texttt{Bilby} \cite{Ashton2019, Ashton2021},
\texttt{PyCBC Inference} \cite{Biwer2019},
\texttt{bajes} \cite{Breschi2021},
\texttt{QuickCBC} \cite{Cornish2021quickcbc},
\texttt{cogwheel} \cite{Roulet2022},
\texttt{simple-pe} \cite{Fairhurst2023},
\texttt{VARAHA} \cite{Tiwari2023}
and
\texttt{jim} \cite{Wong2023}.

\subsection{Likelihood acceleration}
\label{sec:likelihood_acceleration}

We now discuss methods to decrease the computational cost of each likelihood evaluation.
Each evaluation involves generating a frequency-domain waveform $h$ and computing the inner products with the data $\langle d \mid h \rangle$ and with itself $\langle h \mid h \rangle$.
Spelling out the log-likelihood Equation~\ref{eq:gaussian_likelihood}, the following integrals need to be computed:
\begin{equation} \label{eq:likelihood_integral}
    \log \likelihood
    = \Re\, 4 \int \rmd f \frac{d(f) h^*(f; \bm \theta)}{S(f)}
      - \frac 12 4 \int \rmd f \frac{|h(f; \bm \theta)|^2}{S(f)}.
\end{equation}
The straightforward approach is to replace these integrals by a Riemann sum.
Unfortunately, the integrands in Equation~\ref{eq:likelihood_integral} are highly oscillatory functions of the frequency, so we should evaluate them on a large number of frequencies in order to resolve their features.
The maximum frequency must contain the support of the whitened template $h(f) / \sqrt{S(f)}$ (usually several hundred Hz), while the frequency resolution is determined by the duration of the signal and of the noise correlations ($\Delta f \lesssim 1/\tau$, where $\tau$ is the longest of these timescales).
Due to narrow-band spectral lines, noise correlations typically span several seconds, while, depending on the mass of the system, the signal duration can be between a fraction of a second for heavy binary black holes to several minutes for light neutron stars.

Next we review methods to rearrange Equation~\ref{eq:likelihood_integral} so that the integrals can still be evaluated accurately at a coarser frequency resolution.
Importantly, these algorithms require evaluating the waveform at an arbitrary set of frequencies---the speed-up comes from the reduced number of evaluations.
Phenomenological waveform approximants and analytic post-Newtonian models have closed-form expressions and are well suited for this.
On the other hand, time-domain models such as the effective one-body family or numerical relativity surrogates do not benefit from evaluating the waveform at fewer frequencies.

\subsubsection{Heterodyned likelihood / Relative binning}
\label{sec:heterodyne}
The \textit{heterodyne} \cite{Cornish2010}, or \textit{relative binning} \cite{Zackay2018} algorithm allows to integrate oscillatory functions---such as encountered in likelihood computation---at low resolution, in the vicinity of a reference integrand function.
The key observation is that waveforms supported by the posterior should be similar to each other, since they all fit the data well.
We can therefore pick one reference waveform $h_0(f)$, e.g.\ by maximizing the likelihood, and describe any similar $h(f; \bm \theta)$ in terms of a smooth departure from it.
The reference waveform is computed at full resolution but only once, while the smooth corrections are computed for multiple $\bm \theta$ but at low frequency resolution.
The frequency resolution required, and thereby the computational cost, turn out to be independent of the duration of the signal \cite{Zackay2018}.
Several variants of the algorithm exist \cite{Cornish2010, Zackay2018, Leslie2021}; borrowing heavily from these ideas, here we present our own formulation.

Let us decompose the strain waveform into harmonics with smooth phase evolution,
\begin{equation}    
    h(f) = \sum_m h_m (f),
\end{equation}
where each mode $h_m$ is obtained by summing the co-precessing frame modes over all $\ell$ of interest---by Equation~\ref{eq:hm_phase}, modes with the same $m$ have a similar phase evolution.
We may then express Equation~\ref{eq:likelihood_integral} ``mode by mode'' as
\begin{equation} \label{eq:likelihood_integral_hm}
\begin{split}
    \log \likelihood
    =& \sum_m \Re\, 4 \int \rmd f \frac{d(f)}{S(f)} \cdot h_m^*(f; \bm \theta)
    - \frac 12 \sum_{m, m'} 4 \int \rmd f \frac{1}{S(f)} \cdot h_m(f; \bm \theta) h_{m'}^*(f; \bm \theta).
\end{split}
\end{equation}
We have written each integral in Equation~\ref{eq:likelihood_integral_hm} in the form
\begin{equation} \label{eq:I}
    I(\bm \theta) = \int \rmd f \, G(f) \cdot H(f; \bm \theta).
\end{equation}
Our assumption is that, while both $G$ and $H$ may be highly oscillatory, we are only interested in values of $\bm \theta$ for which $H(f; \bm \theta)$ is close to some known $H_0(f)$ in the sense that the ratio $H(f; \bm \theta) / H_0(f)$ is smooth (this requirement motivates the mode-by-mode decomposition \cite{Leslie2021}).
Concretely, we approximate this ratio by a spline $s(f; \bm \theta)$ interpolating it at a coarsely spaced set of $\ncoarse$ nodes $\{f_i\}$:
\begin{equation}\label{eq:spline_approx}
    \frac{H(f; \bm \theta)}{H_0(f)} \approx s(f; \bm \theta),
\end{equation}
with exact equality at the nodes.
It is convenient to write $s(f; \bm \theta)$ as a combination of $\ncoarse$ basis splines $\{s_i(f)\}$, defined by
\begin{equation}
    s_i(f_j) = \delta_{ij} \qquad \forall i, j
\end{equation}
at the nodes $\{f_j\}$, where $\delta_{ij}$ is the Kronecker delta.
Indeed, in this basis the spline coefficients are simply the values to interpolate:
\begin{equation}\label{eq:spline_coeffs}
    s(f; \bm \theta) = \sum_{i=1}^\ncoarse \frac{H(f_i; \bm \theta)}{H_0(f_i)} s_i(f).
\end{equation}
Crucially, Equation~\ref{eq:spline_coeffs} separates the (fine) frequency dependence from the parameter dependence.
We assume that we can evaluate these coefficients directly without the need to compute a full waveform.
Using Equations~\ref{eq:spline_approx} and \ref{eq:spline_coeffs}, we can rewrite Equation~\ref{eq:I} as
\begin{equation} \label{eq:I_rb}
    \begin{split}
        I(\bm \theta)
        = \int \rmd f \, G(f) H_0(f) \frac{H(f; \bm \theta)}{H_0(f)}
        \approx \int \rmd f \, G(f) H_0(f) s(f; \bm \theta)
        = \sum_{i} w_i H(f_i; \bm \theta)
    \end{split}
\end{equation}
where the parameter-independent weights
\begin{align}
    w_i &= \frac{1}{H_0(f_i)}\int \rmd f \, G(f) H_0(f) s_i(f) \label{eq:rb_weights}
\end{align}
are computed ahead of time via a Riemann sum.

Applying Equations~\ref{eq:I_rb} and \ref{eq:rb_weights} to Equation~\ref{eq:likelihood_integral_hm}, the log-likelihood is computed as
\begin{align}
    \log \likelihood
    &\approx \sum_{i, m} \Re \{ u_{i m} h_m (f_i; \bm \theta) \}
     - \frac 12 \sum_{i, m,m'} \Re\{
         v_{imm'}h_m(f_i; \bm \theta) h^*_{m'}(f_i; \bm \theta)\}, \label{eq:rb_likelihood}\\
    u_{im} &= \frac{1}{h^{0*}_m(f_i)} 4\int \rmd f \frac{d(f) h^{0*}_m(f)}{S(f)} s_i(f) \\
    v_{imm'} &= \frac{1}{h^0_m(f_i)h^{0*}_{m'}(f_i)} 4\int \rmd f \frac{h^0_m(f) h^{0*}_{m'}(f)}{S(f)} s_i(f).
\end{align}
\removable{Note that the second sum is symmetric in $m, m'$, which further cuts down computations.}

Heterodyning achieves a two-fold speedup: both the waveform evaluation and the sum over frequencies are performed on the sparse interpolation nodes $\{f_i\}$, whereas in the Riemann sum approach the two operations are performed at full resolution.

The choice of coarse frequencies $\{f_i\}$ is made with the heuristic of keeping the integration errors around each $f_i$ bounded and similar to each other.
Several algorithms have been used: based on the functional form of the post-Newtonian expansion \cite{Zackay2018} or a numerical estimate of the Fisher matrix \cite{Cornish2021}, using adaptive refinement via bisection \cite{Leslie2021}, or progressively and monotonically adding frequencies to the set \cite{Narola2023}.
\removable{As a ballpark estimate, a few hundred frequencies are typically needed, taking on the order of a millisecond to evaluate a waveform with the \texttt{IMRPhenomXPHM} approximant.}

\subsubsection{Reduced-order quadrature}

Reduced-order quadrature is another method to decrease the number frequencies at which the waveform, and integration quadrature, are evaluated.
The method shares many similarities with heterodyning/relative binning, but has the advantage that no reference waveform is required, at the expense of a lower efficiency.

The first step is to build a \textit{reduced-order model} of the waveform as a function of physical parameters.
We approximate the waveform by a model of the form
\begin{equation}\label{eq:rom}
    H(f; \bm \theta) \approx \sum_{i=1}^{N_{\rm coarse}} c_i(\bm \theta) e_i(f),
\end{equation}
where, again, $N_{\rm coarse}$ is small compared to the number of frequencies in a Riemman sum.
The functions $\{e_i(f)\}$ form a basis for a vector space of ``waveforms'', while the functions $c_i$ describe the coefficients that yield the specific waveform of interest (or, rather, its projection on this vector space).
Building this decomposition is the challenging part; once this has been done, the frequency integrals appearing in the likelihood are evaluated in much the same way as in Section~\ref{sec:heterodyne}:
\begin{equation}
\begin{split}
    I(\bm \theta) &\approx \sum_i  w_i c_i(\bm \theta),
    \quad \text{where} \quad
    w_i = \int \rmd f \, G(f) e_i(f).
\end{split}
\end{equation}
The weights $w_i$ are computed at the start of the analysis, once the data is known.

The reduced-order model contains no information about the data, so it must be accurate globally.
An important benefit is that the model is built offline, before any data is collected, and reused for many events.
In general, strategies to build the reduced-order model are based on a large collection of example waveforms that tile the target parameter space.
One such strategy is a ``greedy" algorithm: start with a random waveform as the first basis element, then, define each subsequent element as the waveform that is worst described as a linear combination of the current bases up to that point \cite{Antil2013,Canizares2013,Smith2016}.
Another natural way to choose the basis is through a singular value decomposition of the example waveforms \cite{Purrer2014,Cotesta2020,Morras2023}.
The basis is kept as small as possible while guaranteeing that the waveforms are reproduced with sufficient accuracy.

To obtain the coefficients $\{c_i(\bm \theta)\}$, two main strategies exist.
If we have the ability to evaluate the waveform model at arbitrary frequencies, then we can choose $N_{\rm coarse}$ interpolation nodes $\{f_j\}$ and solve for the coefficients $\{c_i\}$ in terms of $\{H(f_j)\}$---again, similarly to Section \ref{sec:heterodyne}.
These interpolation nodes may be chosen with another greedy algorithm: for each $j$, pick $f_j$ as the frequency that maximizes the residual after interpolating $e_j(f)$ at the previous frequency nodes $\{f_{<j}\}$ with a linear combination of the previous basis elements $\{e_{<j}(f)\}$ \cite{Antil2013,Canizares2013,Smith2016}.
More optimal choices for the interpolation nodes are explored in \cite{Morras2023}.
The second strategy is to build fitting formulas for $\{c_i(\bm \theta)\}$ directly, e.g.\ with interpolating splines \cite{Purrer2014,Cotesta2020} or neural networks \cite{Chua2019, Khan2021, Thomas2022}.
This approach is particularly useful if we lack a closed-form expression for the waveform as a function of frequency (see the discussion at the beginning of Section~\ref{sec:likelihood_acceleration}).

\subsubsection{Multibanding}
Gravitational wave inspirals spend most of the time at low frequencies, while high-frequency emission is concentrated towards the end of the inspiral.
Multibanding \cite{Vinciguerra2017, GarciaQuiros2020} exploits this particular morphology to evaluate waveforms on fewer frequencies while preserving accuracy.
Heuristically, the frequency resolution required to accurately represent the waveform is inversely proportional to its duration, and since the high-frequency portion of the signal lasts a shorter time it can be described at coarser frequency resolution.

The frequency domain is partitioned into bands,
\begin{equation}
    h(f) = \sum_b h_b(f),
\end{equation}
and in each band $b$ the waveform is evaluated at a different resolution $\delta f_b \sim 1/\tau_b$.
The amplitude and phase of the waveform can then interpolated to the fine frequency resolution of the original, regular grid \cite{Vinciguerra2017}.
However, in that case the speedup comes exclusively from evaluating the waveform model at fewer frequencies---the frequency integral is still performed at full resolution.
To accelerate the quadrature over frequencies as well, the data can be similarly separated into frequency bands at different resolution \cite{Morisaki2021}.

To choose the bands and resolutions, \cite{Vinciguerra2017} designed a scheme based on the leading post-Newtonian expression for the time to merger as a function of frequency and chirp mass, while \cite{GarciaQuiros2020} used general error bounds for the Taylor series.

Unlike the heterodyne/relative binning method described in Section~\ref{sec:heterodyne}, the computational cost of multibanding does increase with waveform duration, so the achieved speedup is more modest.
One comparative advantage is that a reference waveform is not required, only a target chirp-mass range.

\subsubsection{Interpolation}

Finally, another option for accelerating the likelihood evaluation is to construct an interpolant up-front, which then allows to bypass the expensive computation of the waveform.
Likelihood interpolation is used by the \texttt{RIFT} code \cite{Lange2018}, that employs Gaussian process interpolation to approximate the marginalized likelihood on the space of intrinsic parameters.
Another example is \cite{Pathak2023, Pathak2023b}; in this case they use radial basis function interpolation to compute the matched-filtering timeseries and waveform amplitude, from which the likelihood is calculated.

\subsection{Degeneracy and parametrization}
\label{sec:degeneracy}

An approach complementary to accelerating likelihood evaluations is to ensure that the parameter space is explored in an efficient way, so that less waveform evaluations are needed to achieve convergence.
Generally, stochastic sampling performs best when the distribution is unimodal and approximately Gaussian, which in practice is never the case of gravitational-wave posteriors due to the ubiquitousness of degeneracy.
Degeneracies arise when different values of the parameters produce a similar waveform within measurement uncertainty.
When this happens, the posterior does not have a sharp peak, rather, it stretches along the degenerate directions in parameter space if the degeneracy is continuous, or becomes multimodal if the degeneracy is discrete.
Degeneracies are especially problematic if they produce tight nonlinear correlations involving multiple parameters or if they create disconnected modes, both cases are hard to explore with general-purpose methods and turn out to be relevant to gravitational waves.

We can use to our advantage our analytical knowledge of the posterior, by predicting how degeneracies will appear and devising schemes that deal with them.
\removable{Let us, then, understand how degeneracies arise in gravitational wave posteriors---a comprehensive discussion can be found in \cite{Fairhurst2023}.
Especially for low-mass systems, the inspiral determines the first few coefficients of the post-Newtonian expansion of the waveform phase, Equation~\ref{eq:pn_phase}.
Typically, two independent quantities are measured, which roughly correspond to the chirp mass and a combination of the mass ratio and aligned spins.
Higher modes, when present, may alleviate the mass-ratio--effective-spin degeneracy.
Occasionally, a measurement of precession can constrain the amplitude and phase of the subleading precession harmonic \cite{Fairhurst2023, Fairhurst2020b, Green2021}.
All in all, even for the most favorable sources at most four or five combinations (often less) out of the eight intrinsic parameters get meaningfully constrained.
The remaining seven extrinsic parameters control the amplitude, phase and time of arrival at each detector.\footnote{This is strictly true for quadrupolar, aligned spin waveforms. The viewing angle affects the relative contribution from each harmonic mode and, for precessing systems, the magnitude of the amplitude modulations \cite{Fairhurst2023}. In the majority of cases these effects are subdominant.} 
In principle this allows to measure up to three independent parameters per detector, although in practice less because the two LIGO interferometers have similar orientations.
Thus, extrinsic parameters exhibit significant degeneracy as well, especially when few detectors observe the signal.
What is more, multiple disconnected modes often appear in the sky location, inclination and orbital phase, since those enter the likelihood through oscillatory trigonometric functions that have multivalued inverses \cite{Singer2014}.}

\subsubsection{Continuous degeneracies}

A solution against continuous degeneracies is to redefine the coordinate system so that the parameters become uncorrelated.
The general heuristic is to identify the observables that are constrained by the data and use those as coordinates.

For example, most codes use the chirp mass as coordinate rather than the individual masses \cite{Veitch2015}, which is well constrained since it controls the leading coefficient in the post-Newtonian expansion of the phase.
In more generality, a systematic way of identifying these observables is by constructing the Fisher matrix from derivatives of the log likelihood, which gives an estimate of the inverse covariance \cite{Cutler1994}. 
This is demonstrated by \cite{Morisaki2020, Lee2022}, who use two linear combinations of the first three phase coefficients as coordinates.
They choose these linear combinations as the principal components of the Fisher matrix, as this best removes the correlations.
The reason for using only the two leading principal components is that the subsequent components are in fact more constrained by the prior than by the likelihood (e.g.\ the spins must satisfy $|\chi| < 1$, the symmetric mass ratio $\eta \leq 1/4$), and in that regime Fisher analysis is a poor description of the posterior~\cite{Vallisneri2008}.

For the in-plane spins, the observables are the amplitude and phase of the precession modulations.
These effects are naturally expressed in a coordinate system whose $z$-axis is set by the approximately constant $\bm{\hat J}$, and whose $x$-axis is set using the direction of propagation towards the observer \cite{Farr2014}; other choices of axes produce more correlated posteriors.

Extrinsic parameters are mostly constrained by the observable amplitude, phase and arrival time of the wave at each detector.
The time of arrival at one detector constrains a combination of the geocenter arrival time and the sky location, it can be used directly as a coordinate.
The arrival time at a second detector further confines the source to a ring in the sky, centered on the axis containing the two detectors.
That axis can be used to define a polar coordinate system in which correlations are minimized \cite{RomeroShaw2020, Roulet2022}.
The amplitude of the signal measured at one detector determines a combination of the chirp mass, distance, inclination and sky location---the\textit{ chirp distance}~\cite{Brady2008}---that again is a good coordinate.
Finally, the phase of the wave at the detector constrains a combination of the orbital phase, polarization and sky location that can also be used as a well-measured coordinate \cite{Roulet2022}.

Eccentric systems require two additional parameters, a convenient choice is the initial eccentricity and relativistic anomaly at an orbit-averaged reference frequency \cite{RamosBuades2023}.

When changing coordinate system it is necessary to update the prior, which transforms according to the Jacobian determinant: $p(y) = p(x) \left|{\partial x}/{\partial y} \right|$.
This can limit the flexibility of the reparametrization approach whenever the Jacobian becomes intractable.
The ideas outlined above are also domain-specific, requiring a sound physical intuition about the measurement process.
Some recent approaches effectively change coordinates in an automated way by training normalizing flows on the fly as the sampler progresses in its exploration of the parameter space \cite{Karamanis2022,Wong2023}.
Normalizing flows can be regarded as flexible coordinate changes that simplify the structure of the distribution and have simple Jacobians, we discuss them further in Section~\ref{sec:likelihood_free}.

\subsubsection{Discrete degeneracies}
Approximate discrete symmetries of the posterior may cause multimodality.
Some approximate symmetries are predictable: for example a quadrupole-only of the waveform is invariant under a shift of $\pi$ in the orbital phase.
Similarly, under a simultaneous shift in phase and polarization by $\pi/2$, both the strain $(h_+, h_\times)$ and the antenna response  $(F_+, F_\times)$ change sign, again leaving the measured signal invariant.
For signals dominant in the two LIGO detectors, due to near-alignment of the detectors a flip in $\cos \iota$ or a rotation by $\pi$ in the sky location about the axis containing the two LIGO detectors become approximate symmetries \cite{Singer2014, Roulet2022}.
Alternatively, multimodality can be identified automatically with a preliminary exploration and kernel density estimation \cite{Farr2014b}.

For discrete degeneracies, it is beneficial to connect the modes in some way, so that the sampler can easily explore all of them.
This can be done by designing custom jumps for the sampler to propose \cite{Veitch2015, Cornish2015, Ashton2021}, or by sampling a modified distribution in which the modes of the posterior are folded together \cite{Roulet2022}.
Custom jump proposals can handle a larger variety of jumps, but require modifying the sampler and so far only have been implemented for sampling methods that use Markov Chain Monte Carlo.
In this case it is important to ensure that the jump proposals preserve detailed balance.
Conversely, folding can work with general-purpose samplers, however it requires modifying the sampled distribution and introduces an additional postprocessing step to reconstruct the original posterior.

\subsection{Importance sampling}
\label{sec:importance_sampling}

Importance sampling is a Monte Carlo method that allows to study one distribution while sampling from another \cite{Owen2013}.
The idea is to use samples $\{\bm \theta_i\}$ from a distribution $q$ as if they had been drawn from a different distribution $p$, by assigning weights $w_i = p(\bm \theta_i)/q(\bm \theta_i)$.
We can see how this works by multiplying and dividing by $q(\bm \theta)$ in Equation~\ref{eq:montecarlo_integral}:
\begin{equation}
    \int \rmd \bm \theta\, p(\bm \theta) f(\bm \theta) 
    = \int \rmd \bm \theta\, q(\bm \theta) \frac{p(\bm \theta)}{q(\bm \theta)} f(\bm \theta)
    \approx \frac 1 N \sum_{\bm \theta_i \sim q}^N w_i f(\bm \theta_i).
    \label{eq:importance_sampling}
\end{equation}
Abstracting away the arbitrary function $f$, the weighted samples represent the distribution $p$.
In many applications, neither $p$ nor $q$ need to be normalized.

Importance sampling is subject to pitfalls if the importance distribution $q$ is not chosen adequately.
We can see that the weights are badly conditioned if $q \ll p$ in some region.
If the weights are too imbalanced, we may end up with a few samples dominating the estimation.
A way to quantify this is through the effective sample size
\begin{equation}
    n_{\rm eff} = \frac{\left(\sum_i w_i\right)^2}{\sum_i w_i^2},
\end{equation}
defined in this way so that the variance of the estimator \ref{eq:importance_sampling} is ${\rm Var}_q(f) / n_{\rm eff}$.
A low $n_{\rm eff}$ is a diagnostic that the scheme is failing and more samples, or a better choice of $q$, are needed.
As a general rule, the proposal distribution $q$ should have heavier tails than the nominal distribution $p$ so that the weights do not diverge.

Importance sampling can be advantageous when it is much easier to sample from the importance distribution than from the nominal distribution.
For example, based on the consideration that higher modes usually have a subdominant effect on the posterior but significantly increase the cost of waveform evaluation, \cite{Payne2019} implemented a scheme in which posterior samples are drawn using a quadrupole-only waveform model and reweighted using a model that includes higher modes.
If the sampler requires many evaluations of the posterior per sample produced, this approach can provide a speedup, because the more expensive model only has to be evaluated on the accepted samples.
Using a similar logic, importance sampling was applied by \cite{RomeroShaw2019, RomeroShaw2021} to include eccentricity for samples initially obtained with a quasicircular waveform model, or by \cite{Payne2020} to account for detector calibration uncertainty.
Another example is \texttt{DINGO-IS} \cite{Dax2023}, which reweights samples obtained with machine-learning methods (reviewed in Section~\ref{sec:likelihood_free}). This allows to verify and correct the machine-learned posterior, and moreover to compute the Bayesian evidence, since in this application the proposal is normalized.

Another application in which importance sampling is useful is in population inference, where we are interested in fitting models of the astrophysical distribution to the full catalog of observations.
This requires evaluating integrals of the form \ref{eq:importance_sampling} over a large set of distributions $p$, which we cannot afford to sample every time.

A different regime in which importance sampling is beneficial is when the support of $f \cdot p$ is very different from the support of $p$ in Equation~\ref{eq:importance_sampling}.
In that case we may use a proposal matched to $f \cdot p$ to concentrate the samples around the region where the integrand is important.
This has been used to marginalize over time of arrival and sky location in \cite{Pankow2015, Islam2022}, as we discuss in Section~\ref{sec:marginalization}.

\subsection{Marginalization}
\label{sec:marginalization}

A more elaborate technique is to marginalize the distribution over some of the parameters, thereby reducing the parameter space dimensionality and simplifying the task for the sampler.
We know the dependence of the posterior on extrinsic parameters analytically, which makes them well suited for conditioning and marginalizing without requiring additional calls to the waveform approximant.
Let us partition the parameter space $\bm \theta = (\bm \kappa, \bm \mu)$ into parameters $\bm \kappa$ that are kept and $\bm \mu$ that are marginalized over, the marginal posterior is
\begin{equation}
    p(\bm \kappa \mid d) = \int \rmd \bm \mu \, p(\bm \kappa, \bm \mu \mid d).
\end{equation}
To see how this helps, suppose that we have an efficient way of computing $p(\bm \kappa \mid d)$, and also of sampling from the conditional posterior $p(\bm \mu \mid \bm \kappa, d)$---such that the cost of these operations is comparable to that of evaluating $p(\bm \kappa, \bm \mu \mid d)$, typically dominated by the cost of computing a waveform.
Then, we can divide the problem into first obtaining posterior samples for $\bm \kappa$, and later completing them with $\bm \mu$ taken from the conditional posterior to undo the marginalization.
I.e., $\{\bm \theta_i\} = \{(\bm \kappa_i, \bm \mu_i)\}$ with $\bm \kappa_i \sim p(\bm \kappa \mid d)$ and $\bm \mu_i \sim p(\bm \mu \mid \bm \kappa_i, d)$.
The advantage over sampling the full posterior directly is that the $\bm \kappa$-space is lower dimensional, and possibly better behaved if the $\bm \mu$ parameters exhibit complicated structure.

The \texttt{BAYESTAR} pipeline first used this approach to obtain a low-latency posterior for the source location, conditioned on intrinsic parameters and marginalized over the time and orientation under certain approximations, to enable followup of transient electromagnetic counterparts \cite{Singer2016}.
\cite{Pankow2015} developed a scheme to marginalize all extrinsic parameters by Monte Carlo integration while gridding over the intrinsic, subsequently implemented in the software \texttt{RIFT} \cite{Lange2018}.
\cite{Islam2022} implemented a fast algorithm to perform the extrinsic marginalization for quadrupolar, aligned-spin sources, which \cite{Roulet2024b} generalized to signals with precession and higher modes.
Beyond this, several parameter estimation codes allow to marginalize a subset of the extrinsic parameters, such as orbital phase, distance, time and/or polarization.

One very general integration method is Monte Carlo via importance sampling \cite{Pankow2015, Islam2022}, overviewed in Section~\ref{sec:importance_sampling}.
One draws multiple samples of extrinsic parameters, evaluates their likelihood and estimates the marginalized posterior using Equation~\ref{eq:importance_sampling}.
The challenge is to find a proposal distribution that is at the same time feasible to sample and well matched to the posterior.
To inform time and sky location proposals, \cite{Pankow2015} perform adaptive Monte Carlo, while \cite{Islam2022, Roulet2024b} use estimates of the arrival time at each detector.
Monte Carlo naturally allows to sample the conditional posterior, by drawing from the same weighted samples.

Due to their particular functional form, some parameters are easier to marginalize. 
Integration over orbital phase $\phi$ can be done analytically for quadrupolar signals, that satisfy $h(\phi) = \rme^{\rmi 2 \phi} h(\phi{=}0)$.
Then, $\langle h \mid h \rangle$ is independent of $\phi$ and $\langle d \mid h \rangle = \big|( d \mid h )\big| \cos[2(\phi - \phi_0)]$, where the $(d\mid h)$ inner product is defined as in Equation~\ref{eq:innerproduct} except without taking the real part, and $\phi_0$ is a constant. The marginalized posterior becomes
\begin{equation}\label{eq:phase_marginalization}
    \int_0^{2\pi} \frac{\rmd \phi}{2\pi} \exp\left(
        \big|( d \mid h )\big| \cos[2(\phi - \phi_0)]
        - \frac 12 \langle h \mid h \rangle\right)
    = \rme^{-\langle h \mid h \rangle / 2}
        I_0\left(\big|( d \mid h )\big|\right),
\end{equation}
where $I_0$ is a modified Bessel function of the first kind \cite{Veitch2015}.
To undo the marginalization, given the other parameters we can sample $\phi$ by noting that the conditional posterior (the integrand in Equation~\ref{eq:phase_marginalization}) follows a von Mises distribution for $2\phi$, for which efficient routines exist.
For waveforms with higher modes there is no analytical solution, so a different integration method like a quadrature may be used \cite{Roulet2024b}. 

To marginalize the distance, we use that $h(D) = h_1/D$ with $h_1=h(D{=}1)$.
Hence, $\langle d \mid h \rangle = \langle d \mid h_1 \rangle / D$ and $\langle h \mid h \rangle = \langle h_1 \mid h_1 \rangle / D^2$.
The marginalized posterior
\begin{equation}
    \int_0^\infty \rmd D \, \pi(D) \exp\left(
        \frac{\langle d \mid h_1 \rangle}{D}
        - \frac{\langle h_1 \mid h_1 \rangle}{2 D^2}\right)
\end{equation}
only depends on $\langle d \mid h_1 \rangle$ and $\langle h_1 \mid h_1 \rangle$, so it can be tabulated in advance and evaluated with 2D interpolation in terms of these two quantities.
To better condition the interpolation, in practice it is convenient to use other combinations instead of $\langle d \mid h_1 \rangle$ and $\langle h_1 \mid h_1 \rangle$ as coordinates.
Also, a good quantity to tabulate is the logarithm of the ratio of the marginal likelihood to an analytical approximation, as this has a smaller dynamic range, see \cite{Singer2016}.
Simultaneous marginalization over phase and distance can be done for quadrupolar signals with a similar interpolation, except replacing the likelihood by the phase-marginalized likelihood (Equation~\ref{eq:phase_marginalization}), and changing the inputs to $\big|( d \mid h_1 )\big|$ and $\langle h_1 \mid h_1 \rangle$.

The arrival time can be marginalized by quadrature.
This requires the timeseries of $\langle d \mid h \,\rme^{-\rmi 2 \pi f t}\rangle$ in the vicinity of the event, which can be computed efficiently e.g.\ by a pruned fast Fourier transform or the heterodyne method of Section~\ref{sec:heterodyne}.

\section{LIKELIHOOD-FREE METHODS}
\label{sec:likelihood_free}

The methods outlined in the preceding section all use the same basic framework; they repeatedly query the data's likelihood (and prior) at well-chosen sets of parameters for each event.
In recent years, the development and availability of specialized computing hardware has enabled a different class of approaches to inference, which are collectively called {\em likelihood-free}, or {\em simulation-based} methods \cite{Cranmer2020, Beaumont2019}. 
The various approaches under this umbrella share a few common themes:
\begin{enumerate}
  \item {\em Likelihood:} The functional form of the data's likelihood $\likelihood$ is not directly used; indeed it may not even be known or tractable. Instead, the likelihood is viewed as a distribution for the data that may be easily sampled from (recall the perspective of the likelihood as a model of the data generation process mentioned in Section~\ref{subsec:inference}). 
  \item {\em Posterior:} The traditional approach using methods like MCMC yields samples from the posterior distribution rather than its functional form itself. 
  In contrast, likelihood-free methods typically fit functional forms to the posterior $\posterior$, viewed as a function that takes both $\bm{\theta}$ and $d$ as inputs. 
  It is still advantageous to be able to draw samples from the posterior, e.g., to easily marginalize nuisance parameters and compute weighted integrals via Monte Carlo (see \S\ref{sec:representation}), so functional representations that are easy to sample from are convenient. 
 \item {\em Cost amortization:} The cost of fitting a functional form for the posterior is formidable, and is an extra cost that is absent in the conventional approach. 
 However, this fit is typically performed ahead of time by simulating many datasets (see point 1 above), and once fresh data comes in, the fit can be re-used with negligible computational overhead. 
 In this sense, the costs of performing the fit are {\em amortized} over the many occasions we have to perform parameter inference, to the point where the average computational cost of any single inference can be very low.
\end{enumerate}
For points 1 and 2 above, it is interesting to consider the inversion of perspective in relation to the conventional approach, which samples from the posterior but works with a functional form for the likelihood.

Let us assume we have a flexible parametrization of functions $q_\phi(\bm{\theta} \mid d)$ with some adjustable internal parameters $\phi$, and we want to perform the fit in point 2 above.
For a given $\phi$, we can interpret $q_\phi(\bm\theta \mid d)$ as a function that takes in data $d$ and outputs a probability distribution over $\bm{\theta}$---our approximation to the posterior.
Our task is to optimize $\phi$ to make this approximation as good as we can.
The starting point for this fit is to define a loss function to minimize.
The conventional loss function is the Kullback--Leibler divergence between the true posterior distribution and the fitted form, i.e.,
\begin{align}
  D_{\rm KL}\infdiv{p(\bm{\theta} \vert d)}{q_\phi(\bm{\theta} \vert d)}
  &= \int \rmd\bm{\theta} \,  p(\bm{\theta} \mid d)
      \ln{ \frac{p(\bm{\theta} \mid d)}{q_\phi(\bm{\theta} \mid d)} } \\
  &= \big(\text{term independent of }\bm \phi\big)
      - \int \rmd\bm{\theta} \,  p(\bm{\theta} \mid d) \ln{ q_\phi(\bm{\theta} \mid d)}.
\end{align}
It makes sense to minimize the KL divergence due to its properties that it is nonnegative, and equal to 0 if and only if the two distributions coincide.
We want the fit to perform well for a range of parameters and data and not just one realization, so the actual loss function is averaged over all the possibilities, i.e., 
\begin{align}
  L(\phi) &= - \int \rmd d \, p(d) \int \rmd\bm{\theta} \, 
      p(\bm{\theta} \mid d) \ln{ q_\phi(\bm{\theta} \mid d)} \\
  & = - \int \rmd\bm{\theta} \, \prior \int \rmd d \, \likelihood \ln{ q_\phi(\bm{\theta} \mid d)}. \label{eq:lossintegral}
\end{align}
In the second line, we have rewritten the joint probability $p(\bm{\theta}, d)$ in a way that makes it straightforward to simulate the data generation process: we draw signal parameters $\bm{\theta}_j$ from the prior $\prior$, and then compute the waveforms and add noise to generate synthetic strain data realizations $d_j$, where $j \in \{1,\ldots, N\}$ indexes the samples.
With many simulations, we can evaluate the integral in Equation \ref{eq:lossintegral} using Monte Carlo:
\begin{equation}
\label{eq:lossfunction}
  L(\phi) = - \frac{1}{N} \sum_{\bm{\theta}_j, d_j} \ln{ q_\phi(\bm{\theta}_j \mid d_j)}.
\end{equation}
Given this evaluation of the loss function, one can numerically optimize it to find the best parameters $\phi$ for the function approximation \cite{Chua2020}.
The practical challenge is to propose a family of models $q_\phi$ that is flexible enough to express any posterior that the data may produce, while keeping Equation~\ref{eq:lossfunction} sufficiently amenable to minimization algorithms.

Typically, this challenge is addressed with some use of deep neural networks---functions with an arbitrary number of inputs and outputs, that achieve a high expressiveness by combining thousands of simple, tunable nonlinear components.
A common concern is the extremely large dimensionality of the data $d$; direct `brute force' optimization of the parameters of a network representing the posterior conditioned on such a large number of inputs is not advisable. 
Existing implementations differ in the reduced representation of the data used for the fits, and the exact parametrization of the `model' posterior $q_\phi(\bm{\theta} \mid d)$. 

In the initial work of \cite{Chua2020}, the dimensionality of the data was reduced by projecting it into the lower-dimensional space spanned by basis functions such as the ones used in reduced-order modeling in Equation \ref{eq:rom}. 
The resulting coefficients were used to train simple representations of the marginal posteriors of a subset of the parameters.

Subsequent work that built on this \cite{Green2020, Green2021gw150914, Dax2021} used the technique of {\em normalizing flows} to increase the flexibility of the posterior representation, and achieved successful parameter estimations in the full parameter space for quasi-circular binary mergers using the likelihood-free method. 
A normalizing flow conditioned on the data $d$ (or some reduced representation of it) is a mapping between abstract variables $\bm u$ and the desired physical parameters $\bm{\theta}$, i.e., $\bm{\theta} = f_{\phi, d}(\bm u)$, such that if $\bm u$ are distributed according to some simple distribution of choice (e.g., the standard multivariate normal), under the transformation $f$, the physical variables $\bm{\theta}$ end up being distributed according to the posterior of interest.
A (tractable) normalizing flow is a very powerful representation of the posterior, because it makes both sampling and density evaluation trivial.
Indeed, the flow directly produces samples of $\bm \theta$ given samples of $\bm u$, which can be cheaply generated for simple distributions.
The posterior density in terms of $\bm{\theta}$ is related to the analytical density in the $\bm u$ space by the determinant of the Jacobian of the flow. Mathematically, 
\begin{equation}
  q_\phi(\bm{\theta} \mid d)
  = p_0\left( \bm u {\equiv} f^{-1}_{\phi, d}(\bm{\theta}) \right)
      \left\vert \frac{\partial \bm u}{\partial \bm{\theta}} \right\vert,
\end{equation}
where $p_0$ is the simple base distribution over $\bm u$ (e.g., multivariate normal). Thus, flows with a known Jacobian determinant $\vert \partial \bm u/\partial \bm{\theta}  \vert$ are desirable.
An elegant way to construct expressive yet tractable normalizing flows is by composing multiple simple flows: $f_{\phi, d}(\bm u) = f_N \circ \cdots \circ f_1(\bm u)$, where each element $f_i$ may have parameters that are the output of a neural network that takes the (reduced) data as inputs.
With this construction, the Jacobian determinant is the product of the individual ones. 

Another successful implementation \cite{Gabbard2021} used a variational autoencoder \cite{Kingma2022}, which finds an appropriate reduction of the data as an ingredient in the construction of the model posterior $q_\phi(\bm{\theta} \mid d)$ (as opposed to the coefficients of the reduced-order basis used in \cite{Chua2020}).

\removable{Likelihood-free inference also works for inferring only a subset of the physical parameters; in that case we simply ignore the nuisance parameters after the training data is generated.
In contrast, within likelihood-based inference one must implement an explicit marginalization of the model, as discussed in Section \ref{sec:marginalization}.}

Finally, it has been noted that likelihood-free inference does not require making assumptions about Gaussianity or stationarity of the noise, so long as we are able to simulate it.
One way to simulate detector noise empirically under weak assumptions is to train the model on actual data with synthetic signals injected \cite{Dax2021}.

Codes for likelihood-free inference of gravitational wave sources include
\texttt{PERCIVAL}~\cite{Chua2020},
\texttt{VItamin}~\cite{Gabbard2021}
and
\texttt{DINGO}~\cite{Dax2021}.

\section{CONCLUSION}
\label{sec:conclusion}

Virtually all implications of gravitational wave data for physics and astrophysics\removable{---ranging from multimessenger astronomy to evolution of massive stellar binaries or tests of theories of gravity---}rely on parameter estimation.
There is ample opportunity for the broad astrophysics community to get involved in these endeavors, as these data are periodically released (after an initial proprietary period) through a Gravitational Wave Open Science Center (\url{https://gwosc.org}), along with software and other resources \cite{Abbott2023}.

In this article we have reviewed the theoretical foundations and state of the art methods for retrieving parameters of gravitational wave signals.
We introduced the basics of gravitational wave detection, describing how the different parameters (chirp mass, mass ratio, aligned and in-plane spins, extrinsic parameters) influence the qualitative features of the signal (phase evolution, amplitude modulations, higher harmonics).

Equipped with this intuition, we have described an array of methods for achieving robust and efficient parameter estimation.
Within the traditional paradigm of likelihood-based inference, we discussed algorithms for accelerating likelihood evaluation (heterodyne/relative binning, reduced-order quadrature, multibanding, interpolation) and, complementarily, improving convergence by simplifying the distribution (via reparametrization, approximating distributions or marginalization of nuisance parameters).
We have also covered the emerging field of likelihood-free inference, where advancements in hardware infrastructure, as well as algorithms for fitting very flexible parametric distributions, allow to perform the inference based on a large number of examples of synthetic data rather than explicit likelihood evaluations.
While computationally intensive, this training is performed offline, subsequently allowing inference in real time as new data is collected.

Due to space constraints we have not covered nonstationarity and non-Gaussianity.
Literature on gravitational wave data analysis under weaker assumptions includes \cite{Chatziioannou2019, Cornish2020, Talbot2020, Biscoveanu2020, Talbot2021, Zackay2021}.
In principle, likelihood-free inference can also deal with these issues naturally, by training on synthetic injections made on real data.

\section*{DISCLOSURE STATEMENT}
The authors are not aware of any affiliations, memberships, funding, or financial holdings that might be perceived as affecting the objectivity of this review.

\section*{ACKNOWLEDGEMENTS}

We thank Katerina Chatziioannou for insightful comments on the manuscript, and Eliot Finch and Jonathan Thompson for helpful discussion.
JR acknowldeges support from the Sherman Fairchild Foundation.
TV acknowledges support from National Science Foundation grants 2012086 and 2309360, the Alfred P. Sloan Foundation through grant number FG-2023-20470, the Binational Science Foundation through award number 2022136, and the Hellman Family Faculty Fellowship.

\bibliographystyle{ar-style5}
\bibliography{main}

\end{document}